\documentclass[usegraphicx]{mn2e}

\newif\ifAMStwofonts

\newcommand{\kms}{$\hbox{km}\cdot \hbox{s}^{-1}$}
\newcommand{\hmpc}{$h^{-1}\,\hbox{Mpc}$}
\newcommand{\mhmpc}{{\, h^{-1}\rm Mpc}}

\newcommand{\de}{\delta}
\newcommand{\te}{\theta}

\newcommand{\f}{\frac}

\newcommand{\s}{\sigma}

\newcommand{\bfx}{{\bf x}}

\newcommand{\bfy}{{\bf y}}
\newcommand{\bfk}{{\bf k}}
\newcommand{\bfv}{{\bf v}}
\newcommand{\bfq}{{\bf q}}
\newcommand{\bfg}{{\bf g}}

\newcommand{\calO}{{\cal O}}

\newcommand{\calL}{{\cal L}}

\newcommand{\calN}{{\cal N}}

\newcommand{\bc}{\begin{center}}
\newcommand{\be}{\begin{equation}}
\newcommand{\ee}{\end{equation}}
\newcommand{\ec}{\end{center}}
\newcommand{\lan}{\langle}
\newcommand{\ran}{\rangle}
\newcommand{\spose}[1]{\hbox to 0pt{#1\hss}}
\newcommand{\lta}{\mathrel{\spose{\lower 3pt\hbox{$\mathchar"218$}}
 \raise 2.0pt\hbox{$\mathchar"13C$}}}
\newcommand{\gta}{\mathrel{\spose{\lower 3pt\hbox{$\mathchar"218$}}
 \raise 2.0pt\hbox{$\mathchar"13E$}}}

\newcommand{\etal}{{et al.}~}
\newcommand{\err}{r}

\begin{document}

\title[Local Group Velocity Vs.\ Gravity]{Local Group Velocity 
					Versus Gravity: \\
					The Coherence Function}

\author[Chodorowski \& Cieciel\c{a}g]
{Micha{\l} J.\ Chodorowski\thanks{E-mail: michal@camk.edu.pl} and
Pawe{\l} Cieciel\c{a}g\thanks{E-mail: pci@camk.edu.pl}\\ 
Copernicus Astronomical Center, Bartycka 18,
00--716 Warsaw, Poland}

\maketitle
\begin{abstract}
In maximum-likelihood analyses of the Local Group (LG) acceleration,
the object describing nonlinear effects is the coherence function
(CF), i.e.\ the cross-correlation coefficient of the Fourier modes of
the velocity and gravity fields. We study the CF both analytically,
using perturbation theory, and numerically, using a hydrodynamic
code. The dependence of the function on $\Omega_m$ and the shape of
the power spectrum is very weak. The only cosmological parameter that
the CF is strongly sensitive to is the normalization $\s_8$ of the
underlying density field. Perturbative approximation for the function
turns out to be accurate as long as $\s_8$ is smaller than about
$0.3$. For higher normalizations we provide an analytical fit for the
CF as a function of $\s_8$ and the wavevector. The characteristic
decoherence scale which our formula predicts is an order of magnitude
smaller than that determined by Strauss \etal\ This implies that
present likelihood constraints on cosmological parameters from
analyses of the LG acceleration are significantly tighter than
hitherto reported.
\end{abstract}


\section{Introduction}
\label{sec:intro}
The dipole anisotropy of the Cosmic Microwave Background (CMB)
temperature is widely believed to reflect, via the Doppler shift, the
motion of the Local Group (LG) with respect to the CMB rest
frame. When transformed to the barycentre of the LG, this motion is
towards $(l,b) = (276^{\circ} \pm 3^{\circ},30^{\circ} \pm
2^{\circ})$, and of amplitude $v_{\rm LG} = 627 \pm 22$ \kms, as
inferred from the 4-year COBE data (Lineweaver \etal 1996).
Alternative models which assume that the dipole is due to a metric
fluctuation (e.g., Paczy\'nski \& Piran 1990) have problems with
explaining its observed achromaticity and the relative smallness of
the CMB quadrupole.

An additional argument in favour of the kinematic interpretation of
the CMB dipole is its remarkable alignment with the LG gravitational
acceleration (or gravity), inferred from galaxy distribution. The
acceleration on the LG, inferred from the {\em IRAS\/} PSCz survey
points only $\sim 13^\circ$ away from the CMB dipole apex (Schmoldt
\etal 1999; hereafter S99, Rowan-Robinson \etal 2000). The alignment
between the two vectors is expected in the linear regime of
gravitational instability (Peebles 1980), and under the hypothesis of
linear biasing between galaxies and mass. The ratio of the amplitudes
of the velocity and gravity vectors is then a measure of the quantity
$\beta = \Omega_m^{0.6}/b$, where $\Omega_m$ and $b$ are the
cosmological density of nonrelativistic matter and linear bias
parameters, respectively. Therefore, comparisons between the LG
gravity and the CMB dipole can serve not only as a test for the
kinematic origin of the latter but also as a measure of
$\beta$. Combined with other constraints on bias, they may yield an
estimate of $\Omega_m$ itself.

However, the linear estimate of the LG velocity from a particular
redshift survey will in general differ from its true velocity. The
reasons are the finite volume of the survey, shot noise due to
discrete sampling of the galaxy density field, redshift-space
distortions and nonlinear effects. In a proper process of the LG
velocity--gravity comparison, all these effects should be accounted
for. 

A commonly applied method of constraining cosmological parameters by
the LG velocity--gravity comparison is a maximum-likelihood analysis,
elaborated by several authors (especially by Strauss \etal 1992,
hereafter S92; see also Juszkiewicz, Vittorio \& Wyse 1990, Lahav,
Kaiser \& Hoffman 1990, S99). In this approach one maximizes the
likelihood of particular values of cosmological parameters given the
observed values of the LG velocity and gravitational acceleration.
This enables one to constrain $\beta$ and the relative amount of power
on large scales within the framework of a given cosmology. 

The analysis of S92 constrained $\beta$ to lie between $0.4$ and
$0.85$ (1 $\s$). The acceleration on the LG was derived there from the
$1.2$ Jy survey of {\em IRAS\/} galaxies. S99 repeated this analysis,
with the LG gravity inferred from the recently completed {\em IRAS\/}
PSCz catalogue. This catalog contains almost three times more galaxies
than its 1.2 Jy subsample. Still, the errorbars on $\beta$, obtained
by S99 ($\beta = 0.70^{+ 0.35}_{-0.2}$ at 1 $\s$), are not smaller
than those obtained by S92. The volume surveyed is larger, shot noise
is suppressed, but the errors remain big. Why? The authors blame
nonlinear effects.

In nonlinear regime non-local nature of gravity is unveiled and the
local relationship between the acceleration and velocity vectors is
partly spoiled. In other words, the nonlinear velocity--gravity
relation at a given point has scatter. As a result, the precision of
determining $\beta$ by the method described above is fundamentally
limited, regardless how well we can measure the LG gravity (and
velocity).

This argument sounds reasonable. However, if nonlinear effects are so
strong, why is the misalignment angle between the LG gravity and
velocity so small? Doesn't it actually suggest otherwise? This
motivated us to reanalyze nonlinear effects in the LG
velocity--gravity comparison.

In a maximum-likelihood analysis, a proper object describing nonlinear
effects is the coherence function (hereafter CF), i.e. the
cross-correlation coefficient of the Fourier modes of the gravity and
velocity fields (see S92 for details).\footnote{S92 call it the {\em
decoherence} function. We prefer the name `coherence', because higher
values of the function imply higher, not lower, correlation between
velocity and gravity.} S92 devised a formula for the CF, calibrating
it so as to fit the results of N-body simulations of a {\em standard}
CDM cosmology. They adopted this form of the function in all
subsequent analyses, thus ignoring any possible dependence it may have
on the shape of the power spectrum and its normalization. A similar
approach was adopted by S99, who followed `the S92 assumption that the
CF does not change appreciably with the background cosmology'. S99
applied the same form of the function in two different cosmological
models: spatially flat CDM cosmologies with respectively zero and
non-zero ($\Omega_\Lambda = 0.7$) cosmological constant. The adopted
value for the spectral parameter $\Gamma$ was $0.25$ and the spectra
were cluster-normalized, so they had different values of $\sigma_8$
(r.m.s.\ mass fluctuations on the 8 \hmpc\ scale).

By definition, on large enough, linear scales the CF is unity. On
smaller scales we expect it to begin to depart from this value. The
scale of departure marks a characteristic scale at which the non-local
nature of gravity can no longer be ignored. Non-locality of the
fields must be somehow correlated with their non-linearity, because it
does not appear for linear fields. Since in more evolved models the
nonlinearity scale is larger, it is natural to expect the CF to depend
on the normalization of the underlying power spectrum. If this is
indeed the case, then this dependence should be modelled, to be
accounted for in future LG velocity--gravity comparisons. The
dependence on other cosmological parameters should also be
studied. This is the aim of the present paper. It is organized as
follows. In Section~\ref{sec:anal} we calculate the CF
perturbatively. In Section~\ref{sec:num} we describe the numerical
simulations which we use to estimate the CF numerically. Results of
the simulations are presented in Section~\ref{sec:res}. In particular,
we compare the analytical estimates with the numerical estimates of
the CF. In Section~\ref{sec:probab} we show that the CF significantly
affects the estimation of cosmological parameters in
maximum-likelihood analyses of the LG acceleration. Moreover, we
constrain the CF in an alternative way, adopted by S92. Summary and
conclusions are in Section~\ref{sec:conc}.

\section{Analytical calculations}
\label{sec:anal}
The CF is defined as 

\be
C(\bfk) = \f{\lan \bfg_\bfk \cdot \bfv_\bfk^\star \ran}{\lan
|\bfg_\bfk|^2 \ran^{1/2} \lan |\bfv_\bfk|^2 \ran^{1/2}} \,,
\label{eq:dec_def} 
\ee
where $\bfg_\bfk$ and $\bfv_\bfk$ are the Fourier components of the
gravity and velocity fields, and $\lan \ldots \ran$ means the ensemble
averaging. (Note that the definition of S92 lacks the complex
conjugate sign.) The function can be interpreted as the
cross-correlation coefficient of the Fourier modes of the gravity and
velocity fields. It is important for the LG gravity--velocity
comparisons, because it appears in the likelihood function for the LG
velocity and acceleration (see Section~\ref{sec:probab}). As argued in
the Introduction, the CF is also interesting in its own right -- it
carries information about non-locality, and indirectly about
non-linearity, of the fields. In this section we calculate the CF for
the fields which are {\em weakly} non-linear.

By definition, $C(\bfk)$ is invariant to scaling of $\bfg_\bfk$ and
$\bfv_\bfk$ by an arbitrary constant. We are then free to choose the
fields scaled so as to fulfill the following equations:
\begin{eqnarray}
\nabla \cdot \bfg &=& \de\,, \\
\nabla \cdot \bfv &=& \te\,. 
\label{eq:diadys}
\end{eqnarray}
Here, $\de$ is the mass density contrast and $\te$ is the velocity
divergence, scaled in such a way that in the linear regime $\te =
\de$. 

The gravity field is strictly irrotational. Hence

\be
\bfg_\bfk = \f{i \bfk}{k^2} \de_\bfk \,,
\label{eq:gravity}
\ee
where $\de_\bfk$ is the Fourier transform of the density contrast. Due
to Kelvin's circulation theorem, the cosmic velocity field is
vorticity-free as long as there is no shell crossing. Since
appreciable shell crossing does not occur for weakly nonlinear fields,
we have

\be
\bfv_\bfk = \f{i \bfk}{k^2} \te_\bfk \,,
\label{eq:velocity}
\ee
where $\te_\bfk$ is the Fourier transform of the velocity divergence.
Using equations~(\ref{eq:gravity}) and~(\ref{eq:velocity}) we obtain 
  
\be
C(\bfk) = \f{\lan \de_\bfk {\te_\bfk}^{\!\!\star} \ran}{\lan \de_\bfk
{\de_\bfk}^{\!\!\star} \ran^{1/2} 
\lan \te_\bfk {\te_\bfk}^{\!\!\star}\ran^{1/2}} \,.
\label{eq:dec_scal} 
\ee

We assume here that the density and velocity divergence fields are
homogeneous and isotropic random fields. For such fields, the
CF depends only on the magnitude of the
wavevector. Furthermore, the density and velocity divergence are real
fields, what implies that ${\de_\bfk}^{\!\!\star} = \de_{- \bfk}$, and
similarly for the velocity divergence. All this implies that the
CF is a real function, as follows. Namely, we have
$C(\bfk) = C(-\bfk)$, hence $\lan \de_\bfk {\te_\bfk}^{\!\!\star}\ran
= \lan \de_{-\bfk} {\te_{-\bfk}}^{\!\!\star}\ran = \lan
{\de_{\bfk}}^{\!\!\star} \te_\bfk\ran = {\lan
\de_\bfk {\te_\bfk}^{\!\!\star}\ran}^\star$. Therefore, $C(k) =
[C(k)]^\star$, that is $C(k)$ is real. We may then cast it
to an explicitly real form: 

\be 
C(k) = \f{\lan \de_\bfk {\te_\bfk}^{\!\!\star} \ran + \lan
{\de_{\bfk}}^{\!\!\star} \te_\bfk\ran}{2 \lan \de_\bfk
{\de_\bfk}^{\!\!\star} \ran^{1/2}
\lan \te_\bfk {\te_\bfk}^{\!\!\star}\ran^{1/2}} \,.
\label{eq:dec_real} 
\ee
This formula is a good approximation to the CF as long as the velocity
field has negligible vorticity. N-body simulations (Bertschinger \&
Dekel 1989, Mancinelli \etal 1994, Pichon \& Bernardeau 1999) show
that even in the case of fully nonlinear fields, the generated
vorticity is small.


We now expand $\de_\bfk$ and $\te_\bfk$ in perturbative series: $\de_\bfk =
\de_\bfk^{(1)} + \de_\bfk^{(2)} + \de_\bfk^{(3)} + \ldots$, and similarly 
$\te_\bfk = \te_\bfk^{(1)} + \te_\bfk^{(2)} + \te_\bfk^{(3)} +
\ldots$. Since in the linear regime $\te = \de$, we have
$\te_\bfk^{(1)} = \de_\bfk^{(1)}$. We assume here that the initial
(linear) density fluctuation field is a Gaussian random field. For
such a field, all odd-order moments of the density contrast and of the
velocity divergence vanish.  Then, $\lan \de_\bfk
{\te_\bfk}^{\!\!\star} \ran =
\bigl\lan\de_\bfk^{(1)} {\te_\bfk^{(1)}}^{\star} \bigr\ran + 
\bigl\lan\de_\bfk^{(1)} {\te_\bfk^{(3)}}^{\star} \bigr\ran + 
\bigl\lan\de_\bfk^{(3)} {\te_\bfk^{(1)}}^{\star} \bigr\ran + 
\bigl\lan\de_\bfk^{(2)} {\te_\bfk^{(2)}}^{\star} \bigr\ran + 
\calO\left(\s^6\right)$, where $\s^2 \equiv \lan\de^2\ran$, and
similarly for other terms appearing in formula~(\ref{eq:dec_real}).
Up to the leading order corrective term, this expansion yields

\be
C(k) = 1 - \f{\Bigl\lan \bigl|\de_\bfk^{(2)} - \te_\bfk^{(2)}\bigr|^2 
\Bigr\ran}{2 (2 \pi)^3 P(k)} \,.
\label{eq:dec_pert}
\ee
Here, $P(k)$ is the {\em linear} power spectrum, defined as $(2
\pi)^{-3} \bigl\lan \de_\bfk^{(1)} {\de_\bfk^{(1)}}^\star\bigr\ran$.
 
The above formula is somewhat similar to that describing weakly
nonlinear corrections to the evolution of the power spectrum (Makino
\etal 1992, Jain \& Bertschinger 1994). There is, however, also an
important difference. Namely, all terms of the sort
$\bigl\lan\alpha_\bfk^{(1)} {\beta_\bfk^{(3)}}^{\star} \bigr\ran$,
where $\alpha$ and $\beta$ stand for either $\de$ or $\te$, have
remarkably cancelled out. In other words, unlike the weakly nonlinear
power spectrum, the weakly nonlinear CF is constructed solely from
second-order terms.

To proceed further, we need the forms of $\de_\bfk^{(2)}$ and
$\te_\bfk^{(2)}$. Second-order solutions for the density contrast and
(scaled) velocity divergence have been shown to depend extremely
weakly on $\Omega_m$ and $\Omega_\Lambda$ (Bouchet \etal 1992, Bouchet
\etal 1995, Bernardeau \etal 1995). This is also true for higher
orders (see App.~B.3 of Scoccimarro \etal 1998). Here we neglect the
weak $\Omega$-dependence. Then (Goroff \etal 1986)

\be
\de_\bfk^{(2)} = \!\int \f{{\rm d}^3 k_1 {\rm d}^3 k_2}{(2\pi)^3}
\de_D(\bfk_1+\bfk_2 - \bfk) J^{(2)}(\bfk_1,\bfk_2)\, \de_{\bfk_1}^{(1)} 
\de_{\bfk_2}^{(1)} 
\label{eq:de_2}
\ee
and

\be
\te_\bfk^{(2)} = \!\int \f{{\rm d}^3 k_1 {\rm d}^3 k_2}{(2\pi)^3}
\de_D(\bfk_1+\bfk_2 - \bfk) K^{(2)}(\bfk_1,\bfk_2)\, \de_{\bfk_1}^{(1)} 
\de_{\bfk_2}^{(1)},
\label{eq:te_2}
\ee
where $\de_D$ is the Dirac delta,

\be
J^{(2)}(\bfk_1,\bfk_2) = \f{5}{7} + \f{2}{7} \f{(\bfk_1 \cdot
\bfk_2)^2}{k_1^2 k_2^2} + \f{(\bfk_1 \cdot \bfk_2)}{2}
\left(\f{1}{k_1^2} + \f{1}{k_2^2}\right)
\label{eq:J_2}
\ee
and

\be
K^{(2)}(\bfk_1,\bfk_2) = \f{3}{7} + \f{4}{7} \f{(\bfk_1 \cdot
\bfk_2)^2}{k_1^2 k_2^2} + \f{(\bfk_1 \cdot \bfk_2)}{2}
\left(\f{1}{k_1^2} + \f{1}{k_2^2}\right) \!.
\label{eq:K_2}
\ee
Hence we have 

\be 
\de_\bfk^{(2)} - \te_\bfk^{(2)} = \!\int \f{{\rm d}^3
k_1 {\rm d}^3 k_2}{(2\pi)^3} \de_D(\bfk_1+\bfk_2 - \bfk)
L(\bfk_1,\bfk_2)\, \de_{\bfk_1}^{(1)} \de_{\bfk_2}^{(1)} \!,
\label{eq:de-te} 
\ee 
where
 
\be
L(\bfk_1,\bfk_2) = \f{2}{7} \left[1 - \left(\f{\bfk_1 \cdot
\bfk_2}{k_1 k_2}\right)^2 \right] .
\label{eq:L}
\ee

To determine $\bigl\lan \bigl|\de_\bfk^{(2)} - \te_\bfk^{(2)}\bigr|^2 
\bigr\ran$, we need to evaluate the four-point correlations of the
linear density field $\de_{\bfk}^{(1)}$. For a Gaussian random field, 
\begin{eqnarray}
& & \left\lan\de_{\bfk_1}^{(1)} \de_{\bfk_2}^{(1)} \de_{\bfk_3}^{(1)} 
\de_{\bfk_4}^{(1)}\right\ran = \nonumber \\
& &(2 \pi)^6 \de_D(\bfk_1 + \bfk_2) \de_D(\bfk_3 + \bfk_4) 
P(k_1) P(k_3) + \nonumber \\
& &(2 \pi)^6 \de_D(\bfk_1 + \bfk_3) \de_D(\bfk_2 + \bfk_4) 
P(k_1) P(k_2) + \nonumber \\
& &(2 \pi)^6 \de_D(\bfk_1 + \bfk_4) \de_D(\bfk_2 + \bfk_3) 
P(k_1) P(k_2) \,.
\label{eq:four_point}
\end{eqnarray}
Using this property, it is a standard perturbative calculation to show
that

\be
\left\lan \bigl|\de_\bfk^{(2)} - \te_\bfk^{(2)}\bigr|^2 \right\ran =
2 \!\int {\rm d}^3 q P(q) P(|\bfk - \bfq|) L^2(\bfq,\bfk-\bfq) \,,
\label{eq:modulus}
\ee
with $L$ given by equation~(\ref{eq:L}). 

Note that $J^{(2)}$ and $K^{(2)}$ have first-order poles as $k_1 \to
0$ or $k_2 \to 0$ for fixed $\bfk$: $J^{(2)} \sim K^{(2)} \sim (1/2)
\cos\vartheta (k_1/k_2 + k_2/k_1)$, where $\vartheta$ is the angle
between $\bfk_1$ and $\bfk_2$. However, in the
expression~(\ref{eq:de-te}) for $\de_\bfk^{(2)} - \te_\bfk^{(2)}$ they
cancel out: the function $L$ has no poles. This results in significant
simplification of the integrand in equation~(\ref{eq:modulus}).

Using formula~(\ref{eq:dec_pert}) we obtain

\be
C(k) = 1 - P^{-1}(k) \!\int\! \f{{\rm d}^3 q}{(2 \pi)^3} P(q) P(|\bfk -
\bfq|) L^2(\bfq,\bfk-\bfq) \,.
\label{eq:dec_int}
\ee 
Now we write the integral in spherical coordinates $q$, $\vartheta$
and $\phi$, the magnitude, polar angle, and azimuthal angle,
respectively, of the wavevector $\bfq$. Then with the external
wavevector $\bfk$ aligned along the $z$-axis the integral over $\phi$
is trivial and simplifies $\int {\rm d}^3 q$ to the form
$2 \pi \int {\rm d} q q^2 \int {\rm d}\cos\vartheta$.  Furthermore, we
have

\be
L^2(\bfq,\bfk-\bfq) = \f{4}{49} \f{\left(1 - \mu^2\right)^2}{\left[1 +
(q/k)^2 - 2 (q/k) \mu \right]^2} \,,
\label{eq:L^2} 
\ee
where $\mu = \cos\vartheta$. This suggests a change of variables $s 
\equiv q/k$. Performing this finally yields 
\begin{eqnarray}
1 - C(k) \!\!\!\!&=&\!\!\!\! \f{4 k^3}{49 (2\pi)^2 P(k)} 
\int_{k_{min}/k}^{k_{max}/k}
{\rm d}s\,s^2 P(ks) \int_{-1}^{1} {\rm d}\mu \times \nonumber \\ 
\!\!\!\!& &\!\!\!\! P[k(1 + s^2 - 2 s \mu)^{1/2}] \, F(s,\mu) \,, 
\label{eq:dec_fin}
\end{eqnarray}
with 

\be
F(s,\mu) = \f{\left(1 - \mu^2\right)^2}{(1 + s^2 - 2 s\mu)^2} \,.
\label{eq:F}
\ee
The quantities $k_{min}$ and $k_{max}$ are the cutoffs of the power
spectrum. Physically, they are related to the effective depth of the
survey, from which the specific spectrum is extracted, and the
virialization scale, respectively. (For $k > k_{max}$ perturbative
expansion breaks down.) In numerical simulations, $k_{min}$ and
$k_{max}$ are determined by the simulation box size and elemental cell
size, respectively. Details will be given in the next Section.

Given the linear spectrum $P(k)$, equation~(\ref{eq:dec_fin}) provides
the second-order correction to the linear value of the CF. Results for
the PSCz spectrum and for the standard-CDM spectrum are presented in
Section~\ref{sec:res}.

\section{Numerical simulations}
\label{sec:num}
Following Peebles (1987), instead of using a N-body scheme we model
cold dark matter as a pressureless cosmic fluid. Using the Eulerian
code CPPA (Cosmological Pressureless Parabolic Advection, see
Kudlicki, Plewa \& R\'o\.zyczka 1996, Kudlicki \etal 2000, Kudlicki
\etal 2001b) we solve its dynamical equations on a uniform grid fixed 
in comoving coordinates. The main improvements of CPPA over the original
Peebles' code are parabolic density and velocity profiles, variable
timestep, periodic boundary conditions and a flux interchange
procedure, implemented as an approximation to the solution of the
Boltzmann equation.

We chose to use a grid-based code rather than a $N$-body code because
it directly produces a volume-weighted velocity field. This is
important because in the definition of the CF,
equation~(\ref{eq:dec_def}), the velocity field is volume-weighted,
not mass-weighted. Moreover, the field is evenly sampled, which is
convenient for FFT techniques.

The linear velocity depends on the cosmological constant
($\Omega_\Lambda$) very weakly (e.g.\ Lahav \etal 1991); this also
holds for higher orders (see Bouchet \etal 1995, Appendix B.3 of
Scoccimarro \etal 1998 and Nusser \& Colberg 1998). Therefore, it was
a good approximation to assume $\Omega_\Lambda = 0$ in our models. We
have thus studied two zero-$\Lambda$ models with $\Omega_m=1$ and
$\Omega_m=0.3$, assuming Gaussian initial conditions. The parameters
of the runs are given in Table~\ref{tab:runs1}.

\begin{table}
\begin{center}
\caption{Parameters of the runs displayed in Figure~\ref{fig1}. The
acronym \textit{l-l} stands for `low $N^3$, low $k_{\rm Nq}$',
\textit{h-l} stands for `high $N^3$, low $k_{\rm Nq}$', and \textit{h-h}
stands for `high $N^3$, high $k_{\rm Nq}$'. \label{tab:runs1}}
\begin{tabular}{|c|c|c|c|}
\hline
run & grid & box~size~[$h^{-1}\rm Mpc$]& $k_{\rm Nq}~[h\, {\rm Mpc}^{-1}]$ \\
\hline \hline
\textit{l-l}  & $64^3$      &  50    &    4.  \\
\hline
\textit{h-l}  & $128^3$     &  100   &    4.  \\
\hline
\textit{h-h}  & $128^3$     &  50    &    8.  \\
\hline
\end{tabular}
\end{center}
\end{table}


To make the simulated gravitational field as close as possible to that
inferred from the {\em IRAS\/} PSCz survey, the mass power spectrum
that we adopted was that estimated for the PSCz galaxies (Sutherland
\etal 1999):
\begin{eqnarray}
\label{eq:power}
P(k) & = & \frac{Bk}{\lbrace 1+[ak+(bk)^{3/2}+(ck)^2]^{\nu}\rbrace^{2/\nu}} \\
a & = & 6.4/\Gamma \mhmpc\,,~~~b=3.0/\Gamma \mhmpc\,, \nonumber\\
c & = & 1.7/\Gamma \mhmpc\,,~~~\nu=1.13 \,, \nonumber
\end{eqnarray}
with $\Gamma = 0.2$ as best fitted value. The power spectrum employed
in simulations is effectively truncated at both large (corresponding
to the lower cutoff $k_{min}$) and small (corresponding to the upper
cutoff $k_{max}$) scales. Specifically, $k_{min} = 2\pi/L$, where $L$
is the simulation box size, and $k_{max} = k_{\rm Nq} = (N/2)
k_{min}$. Here, $k_{\rm Nq}$ is the so-called Nyquist wavevector and
$N^3$ is the grid size.

To normalize the power spectrum we used the observed local abundance
of galaxy clusters. The present value of $\s_8$, labelled $\s_{8,0}$,
is a function of $\Omega_m$ and for the case of $\Omega_\Lambda = 0$
it is estimated by the relation (Eke, Cole \& Frenk 1996)
\be
\s_{8,0} = (0.52 \pm 0.04) \Omega_m^{-0.46 + 0.10 \Omega_m} \,.
\label{eq:sigma_8}
\ee 
This relation changes only slightly with the shape of the power
spectrum. It is also very similar for the case of non-zero $\Omega_\Lambda$,
flat models ($\Omega_\Lambda = 1 - \Omega_m$). For $\Omega_m = 1$, $\s_{8,0}
\simeq 0.52$, while for $\Omega_m = 0.3$, $\s_{8,0} \simeq 0.87$.

\section{Results}
\label{sec:res}
\begin{figure}
\centerline{\includegraphics[angle=0,scale=0.48]{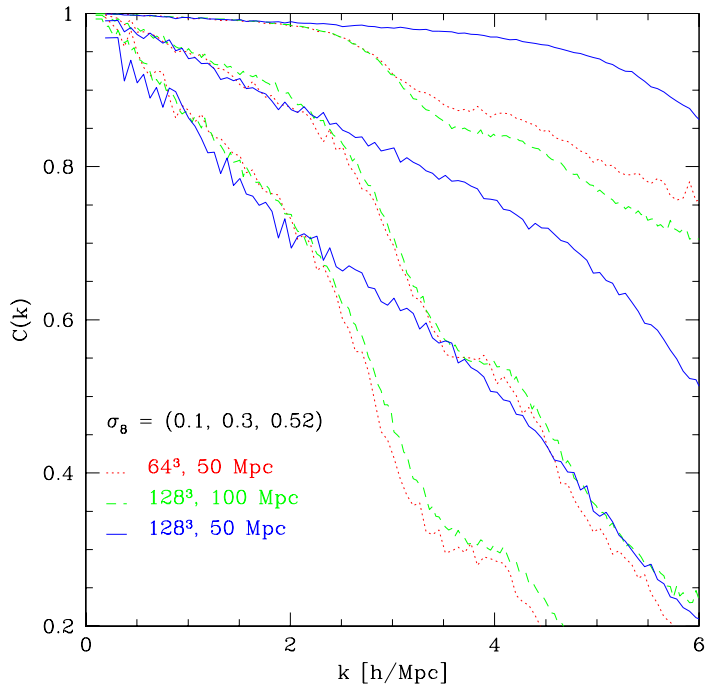}}
\caption{\label{fig1} The CF for three values of $\s_8$,
for three Einstein--de Sitter simulations with different
resolutions. (Their parameters are given in Table~\ref{tab:runs1}.)
Dotted lines show the results of run \textit{l-l}, dashed lines
\textit{h-l}, and solid lines \textit{h-h}. The CF is similar for runs
\textit{l-l} and \textit{h-l}. For $k < 2~h\, {\rm Mpc}^{-1}$, it is
similar for all runs. For each run, the function bends down at $k
\simeq 0.5 k_{\rm Nq}$.}
\end{figure}

Firstly, we tested the dependence of the results on resolution. For
this purpose we performed three simulations of an $\Omega_m=1$ Universe
with the PSCz power spectrum (see Table~\ref{tab:runs1}). The runs
\textit{l-l} and \textit{h-l} have the same spatial resolution but 
different grid resolutions, while the run \textit{h-h} has higher
spatial resolution. In Figure~\ref{fig1}, we present the temporal
evolution of the CF. Specifically, we show it for 3
values of $\s_8$: $0.1$, $0.3$, and $0.52$. As expected, the results
of runs \textit{l-l} and \textit{h-l} do not differ significantly. In
contrast to the run \textit{h-l}, the run \textit{l-l} has no modes
corresponding to the scales greater than $50$ \hmpc. These scales,
however, are well in the linear regime, so $C = 1$ to good accuracy.

In grid simulations, the largest wavevector is the Nyquist wavevector,
corresponding to the Nyquist wavelength, i.e., the smallest wavelength,
of two cells. Its values for the three runs are given in the last
column of Table~\ref{tab:runs1}. Fourier modes with $k > k_{\rm Nq}$
do not have physical meaning. Inspection of Figure~\ref{fig1},
however, shows that the CF bends down already at $k \simeq 0.5 k_{\rm
Nq}$. Such a scaling of the bending point with the Nyquist wavevector
strongly suggests that this is a resolution effect. Given the
similarity of the three curves for $k < 2~h\, {\rm Mpc}^{-1}$, we can
expect the CF to be free of resolution effects up to $k \simeq 0.5
k_{\rm Nq}$, for any grid.

\begin{figure}
\centerline{\includegraphics[angle=0,scale=0.48]{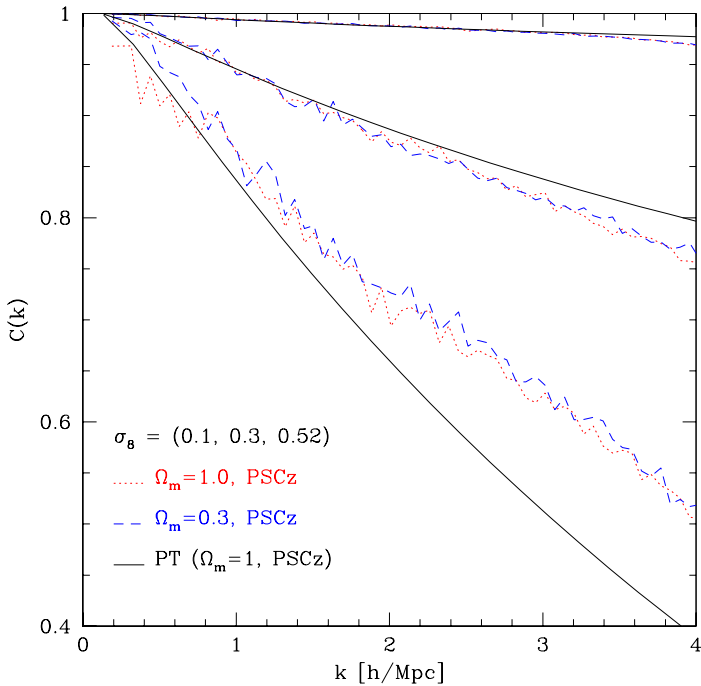}}
\caption{\label{fig2} The CF for simulations $128^3$
with the box-size of $50$ \hmpc. Dotted lines are for $\Omega_m =
1.0$, dashed for $\Omega_m = 0.3$, and solid show predictions of
perturbation theory for $\Omega_m = 1.0$, according to
formula~(\ref{eq:dec_fin}). The CF depends on
$\Omega_m$ very weakly. For $\s_8 \le 0.3$, it is well predicted by the
second-order approximation, except for the highest values of the
wavevector.}
\end{figure}

Next, we test the perturbative approximation for the CF. In
Figure~\ref{fig2}, we show the CF from simulations with the grid
$128^3$ and the box-size of $50$ \hmpc, for wavevectors up to $0.5
k_{\rm Nq} = 4.0~h\, {\rm Mpc}^{-1}$. Dotted lines are for $\Omega_m =
1.0$, dashed for $\Omega_m = 0.3$, while solid show predictions of
perturbation theory for $\Omega_m = 1.0$, according to
formula~(\ref{eq:dec_fin}). (We numerically evaluate the integral in
this formula.)  The CF depends on $\Omega_m$ very weakly. Moreover,
except for the highest $k$-values, it is well predicted by the
second-order approximation, as long as $\s_8$ is smaller than $\sim
0.3$. We have checked that for $\s_8 < 0.3$ the quantity $1 - C(k)$
scales approximately like $\s_8^2$, as predicted by
equation~(\ref{eq:dec_fin}) (the normalization of $P(k)$ is
proportional to $\s_8^2$). For higher values of $\s_8$, the
second-order approximation {\em overestimates} the deviation of $C$
from the linear value, unity. This behaviour is similar to the
nonlinear evolution of the power spectrum, overestimated by
second-order terms (Jain \& Bertschinger 1994).

\begin{figure}
\centerline{\includegraphics[angle=0,scale=0.48]{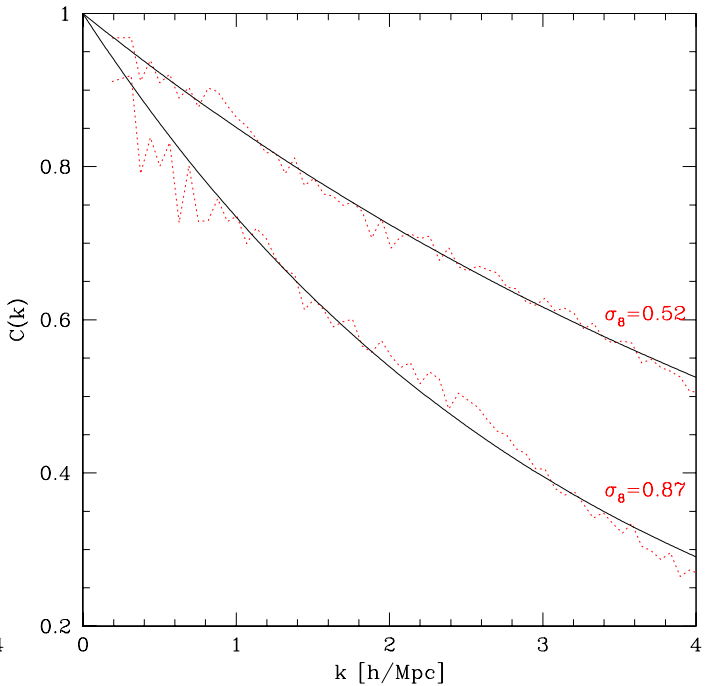}}
\caption{\label{fig3} The coherence function for cluster-normalized
cosmological models. Dotted lines show the function from the numerical
simulation \textit{h-h}, while solid ones are the fits according to
formula~(\ref{eq:func_fit}).}
\end{figure}

To describe the CF for $\s_8 > 0.3$, we analyzed it for 23 output
times of the high-resolution simulation \textit{h-h}, corresponding to
the values of $\sigma_8$ in the range $0.3 < \s_8 < 1.0$. We found
that the $\s_8$-dependence of $C$ can be well fitted as

\be
C(k) = \exp{(-ak)} \,,
\label{eq:func_fit}
\ee
where 

\be
a = \left\{ \begin{array}{ll}
0.757~\sigma_8^2 & \mbox{for $\sigma_8 \le 0.3$} \,, \\
-0.059 + 0.423~\sigma_8 & \mbox{for $0.3 < \sigma_8 \le 1.0$} \,.
\end{array} \right. 
\label{eq:a_fit}
\ee
In Figure~\ref{fig3} we show the CF from the simulation \textit{h-h},
for $\s_8$ given by the cluster normalization. (Having shown very weak
dependence of the function on $\Omega_m$, in this plot we present the
results of Einstein--de Sitter simulations only, which are simpler to
evolve numerically.) In particular, in this figure the dotted line for
$\s_8 = 0.52$ corresponds to the lowest solid line in
Figure~\ref{fig1}. The fits are good; we have checked that for other
values of $\s_8$ they are as good as those shown here.

\begin{figure}
\centerline{\includegraphics[angle=0,scale=0.48]{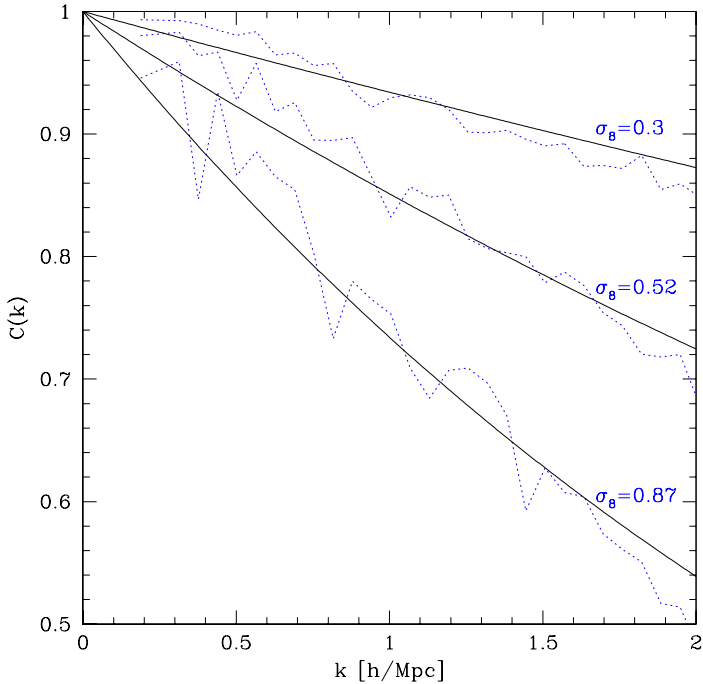}}
\caption{\label{fig3'} The coherence function from a simulation with 
the standard-CDM power spectrum (dotted) versus our
formula~(\ref{eq:func_fit}), obtained for the PSCz spectrum (solid).}
\end{figure}

To test the dependence of the CF on the underlying power spectrum, we
performed a $64^3$ simulation with the standard CDM spectrum ($\Gamma
= 0.5$). The box-size was $50$ \hmpc, so $0.5\, k_{\rm Nq} = 2$. The
results are shown in Figure~\ref{fig3'}. Solid lines are drawn
according to our fit (eq.~\ref{eq:func_fit}), obtained for the PSCz
spectrum, while dotted lines are from the simulation. A good agreement
between them apparent in the figure implies that the dependence of the
CF on the power spectrum is very weak, at least for a CDM-like family
of the spectra.

\begin{figure}
\centerline{\includegraphics[angle=0,scale=0.48]{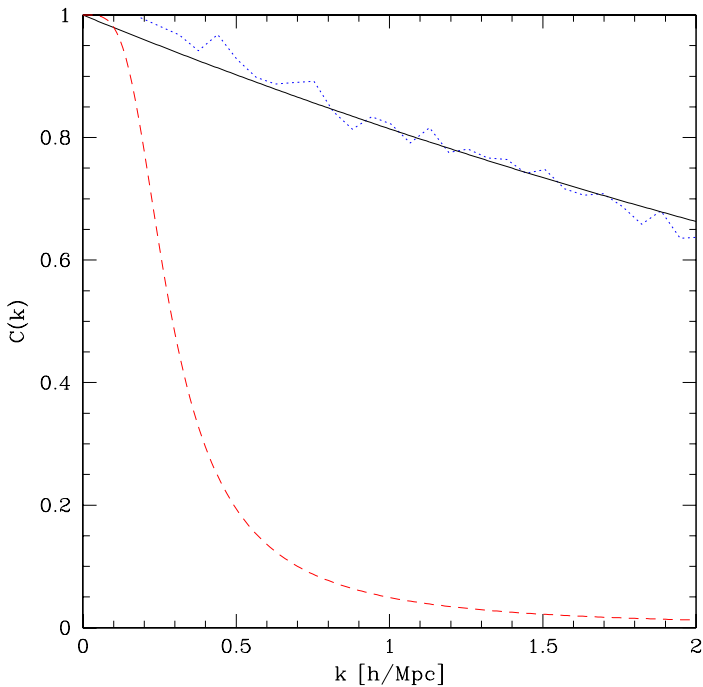}}
\caption{\label{fig4} The coherence function for a standard-CDM
cosmological model with $\s_8 = 0.625$. The dotted line shows the
function from our standard-CDM simulation, while the solid one is the
fit according to formula~(\ref{eq:func_fit}). The dashed line is the
formula~(18) of S92, with $r_c = 4.5$ \hmpc.}
\end{figure}

The CF has been modelled by S92, who calibrated it so as to fit the
results of N-body simulations of a standard CDM cosmology. In
Figure~\ref{fig4} we show S92's prediction for $C$, as well as our
predictions, for the standard CDM power spectrum and $\s_8$
normalization of S92 ($0.625$). The discrepancy of our results with
the formula of S92 is drastic!  Instead of a characteristic
decoherence scale of $4.5$ \hmpc\ (S92), our
formula~(\ref{eq:func_fit}) suggests a fraction of a megaparsec. We
will comment on that in the summary.

\section{From coherence to probability contours}
\label{sec:probab}
The aim of this Section is twofold. Firstly, we show that the CF is
very important in the analyses of the LG acceleration, because,
together with other factors, it determines the relative likelihood of
different cosmological models. Secondly, we show that the probability
distribution for the LG acceleration amplitude and the misalignment
angle, resulting from our estimate of the CF, is consistent with that
obtained from mock {\em IRAS\/} catalogs, constructed by S92. In
contrast, adopting S92's formula for the function results in a
distribution which is too broad. Here we only outline the necessary
formalism; for more details the reader is referred to S92 and
Chodorowski \& Cieciel\c ag (2001).

Let $f(\bfg,\bfv)$ denote the probability density distribution for
observing particular values of the LG gravity and velocity, given some
assumed CDM cosmological model. The model is fully specified by the
value of $\beta$, the power spectrum shape parameter $\Gamma$, and the
normalization $\s_8$. Due to Bayes theorem, it is possible to evaluate
from $f$ the relative likelihood, $\calL$, of different models
($\beta$, $\Gamma$, $\s_8$). The likelihood function is usually
defined as

\be
\calL = f \,.
\label{eq:likelihood}
\ee
Here, the observed values of the Local Group velocity and gravity are
inserted in $f$ and serve as constraints on cosmological parameters.

As a functional form of $f$, S92 and S99 adopt a multivariate
Gaussian. This assumption has support from numerical simulations
(Koffman \etal 1994, Kudlicki \etal 2001a), where the measured
nongaussianity of $\bfg$ and $\bfv$ is small. This is rather natural
to expect since, e.g., gravity is an integral of density over
effectively a large volume, so the central limit theorem can at least
partly be applicable (but see Catelan \& Moscardini 1994). However,
the approximate Gaussianity of $\bfg$ and $\bfv$ by no means implies
that the fields are linear. In contrast, the above considerations
suggest a Gaussian approximation for the form of $f$, but with the
covariance matrix calculated accounting for the nonlinear effects.

After some algebra (see e.g. Juszkiewicz \etal 1990), the joint
distribution for $\bfg$ and $\bfv$ can be cast to the form:

\be
f(\bfg,\bfv) = \f{(1 - \err^2)^{-3/2}}{(2 \pi)^{3} \s_g^{3} \s_v^{3}}
\exp\left[- \f{x^2 + y^2 - 2 \err \mu x y}{2(1 - \err^2)}\right] \,,
\label{eq:dist}
\ee
where $\s_g$ and $\s_v$ denote the r.m.s.\ values of a single spatial
component of gravity and velocity, respectively. From statistical
isotropy, $\s_g^2 = \lan \bfg\cdot\bfg \ran/3$, and $\s_v^2 = \lan
\bfv\cdot\bfv\ran/3$. Next, $(\bfx,\bfy) = (\bfg/\s_g,\bfv/\s_v)$, and
$\mu = \cos\theta$ with $\theta$ being the misalignment angle between
$\bfg$ and $\bfv$. Finally, $\err$ is the cross-correlation
coefficient of $g_i$ with $v_i$, where $g_i$ ($v_i$) denotes an
arbitrary spatial component of $\bfg$ ($\bfv$). From isotropy,

\be
\err = \f{\lan \bfg \cdot \bfv \ran}{\lan g^2 \ran^{1/2} \lan v^2
\ran^{1/2}} \,.
\label{eq:rho}
\ee
Also from isotropy, 
\be
\lan x_i y_j \ran = \err \, \de_{ij} \,,
\label{eq:cross}
\ee
where $\de_{ij}$ denotes the Kronecker delta. In other words, there
are no cross-correlations between different spatial components. 

In the limit of linear fluctuations and with perfect sampling of the
density field, the distribution~(\ref{eq:dist}) reduces to $\de_D(\bfx
- \bfy) \calN_3(\bfv)$, where $\calN_3$ denotes a tri-variate normal
distribution. In a real world, there are nonlinear effects (NL), as
well as finite-volume effects (FV), which make the cross-correlation
coefficient deviate from unity. In a related paper (Chodorowski \&
Cieciel\c ag 2001) we will show that $\err$ can be approximately
written as

\be
\err = \err_{\rm NL} \cdot \err_{\rm FV} \,,
\label{eq:rho_fac}
\ee
i.e. that the contributions to $\err$ from nonlinear effects separate
from those from finite volume. In the present paper we study nonlinear
effects. The cross-correlation coefficient due to them is

\be
\err_{\rm NL} = \f{\int_0^\infty C(k) W_g(k) W_v(k) P(k)\, {\rm
d}k}{\int_0^\infty W_g(k) W_v(k) P(k)\, {\rm d}k} \,.
\label{eq:rho_ne}
\ee
Here $W_g$ and $W_v$ are the observational filters of $\bfg$ and
$\bfv$ (for details see S92). Thus, nonlinear effects enter into the
correlation coefficient of the joint distribution function via the
CF. In the linear regime $C = 1$ and, if we neglect other effects,
$\err = 1$. Since the distribution function is directly related to the
likelihood of different world models (eq.~\ref{eq:likelihood}), in
analyses of the LG acceleration, the CF affects the estimation of
cosmological parameters.

\begin{figure}
\centerline{\includegraphics[angle=0,scale=0.48]{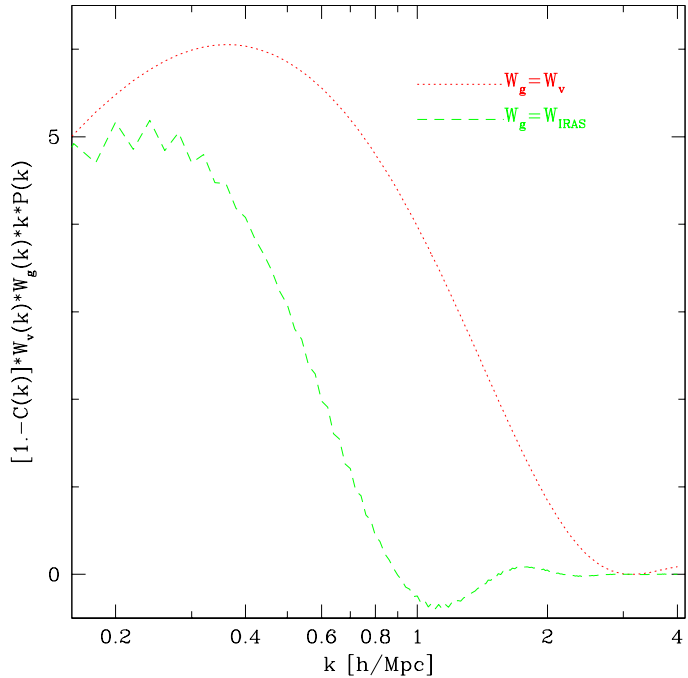}}
\caption{\label{fig:k_range} The integrand in the numerator of the
right-hand side of equation~(\ref{eq:1-rho_ne}) as a function of the
wavevector. Since the $k$-axis is logarithmic, the ordinate is
multiplied by and extra power of $k$, so equal areas under the
function correspond to equal contributions to the deviation of the
correlation coefficient from unity. Units of the ordinate axis are
arbitrary.}
\end{figure}

In equation~(\ref{eq:rho_ne}) the CF is multiplied by the windows
through which the gravity and velocity of the LG are measured. 
Therefore, smoothing effectively filters out the high-$k$
tail of the coherence function. It is instructive to write

\be
1 - \err_{\rm NL} = \f{\int_0^\infty \left[1 - C(k)\right] W_g(k)
W_v(k) P(k)\, {\rm d}k}{\int_0^\infty W_g(k) W_v(k) P(k)\, {\rm d}k}
\,;
\label{eq:1-rho_ne}
\ee
the stronger the deviation of $C$ from unity so is the deviation of
$\err_{\rm NL}$. In Figure~\ref{fig:k_range} we plot the integrand $[1
- C(k)] W_g(k) W_v(k) P(k)$ as a function of $k$.  (Strictly speaking,
we plot the function $[1 - C(k)] W_g(k) W_v(k) k P(k)$, because the
$k$-axis is logarithmic.) Here $C(k)$ is given by our fit
(eq.~\ref{eq:func_fit}), $P(k)$ is the spectrum of the PSCz galaxies,
and the velocity window is that introduced by S92, with a small-scale
cutoff, $r_{min} = 1$ \hmpc, to reflect the finite size of the LG. The
gravity window is either the standard {\em IRAS\/} window with a
small-scale smoothing $r_s = 5$ \hmpc\ (S92), or equal to the velocity
window. The first case corresponds to the present situation, where the
inferred LG gravity is commonly smoothed with the standard {\em
IRAS\/} window, and results in the dashed curve. The second case
describes an ideal situation, where the galaxy distribution around the
LG is sampled so densely that there is no need to smooth its gravity
beyond the size of the LG. That case results in the dotted curve. We
see from the figure that at present it is sufficient to know the
behaviour of the CF up to at most $k = 2~h\, {\rm Mpc}^{-1}$, and it
will never become necessary to know it for $k > 4~h\, {\rm
Mpc}^{-1}$. This is why we have set up the resolution of the
\textit{h-h} simulation in such a way that $0.5\, k_{\rm Nq} = 4~h\,
{\rm Mpc}^{-1}$ (Section~\ref{sec:res}).

In contrast to our approach, S92 did not determine the CF from its
definition. Instead, using standard CDM N-body simulations, they
created mock {\em IRAS\/} catalogs for `observers' (N-body points)
selected with similar properties to those of the LG (primarily the
velocity). For each of these observers, S92 computed the {\em IRAS\/}
acceleration of the LG, as the filter $W_g$ using the so-called
standard {\em IRAS} window. They then matched the resulting
distribution for the amplitude of $\bfg$ and the misalignment angle
with the probability contours resulting from a formula following from
equation~(\ref{eq:dist}). Since the LG velocity was considered by S92
as a constraint, we need the conditional probability density function
$f(\bfg|\bfv)$. It readily results from formula~(\ref{eq:dist}):

\be
f(\bfg|\bfv) = (2 \pi)^{-3/2} \s_g^{-3} (1 - \err^2)^{-3/2} 
\exp\left[- \f{(\bfx - \err \bfy)^2}{2(1 - \err^2)}\right] 
\label{eq:dist_cond}
\ee
(Juszkiewicz \etal 1990). The distribution for $g$ and $\theta$
results from the above formula by multiplying it by $2 \pi g^2
\sin\theta$. 

\begin{figure}
\centerline{\includegraphics[angle=0,scale=0.51]{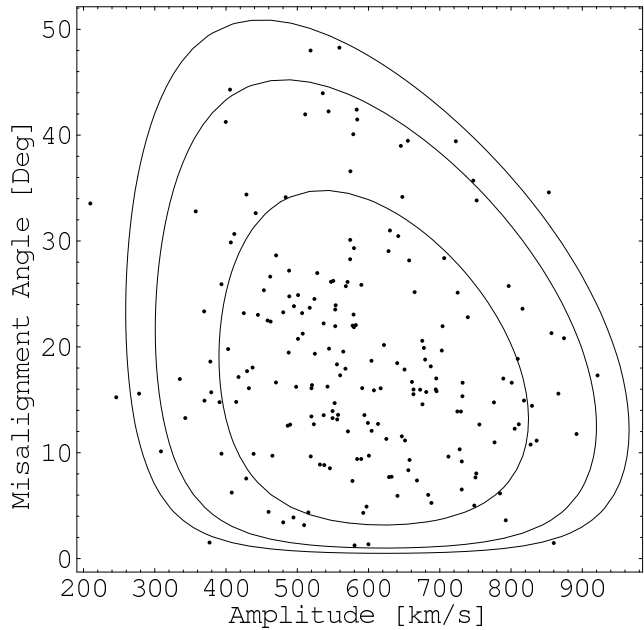}}
\caption{\label{fig:new} Contours of probability density as a function of
amplitude of the LG gravity and the misalignment angle. The LG
velocity (CMB dipole) is used here as a constraint. The contours
correspond to the probability levels of 68\%, 90\%, and 95\%. The
coherence function is given by our formula~(\ref{eq:func_fit}). The
scattered points in this Figure and in Figure~\ref{fig:old} are S92's
values from simulations.}
\end{figure}

\begin{figure}
\centerline{\includegraphics[angle=0,scale=0.511]{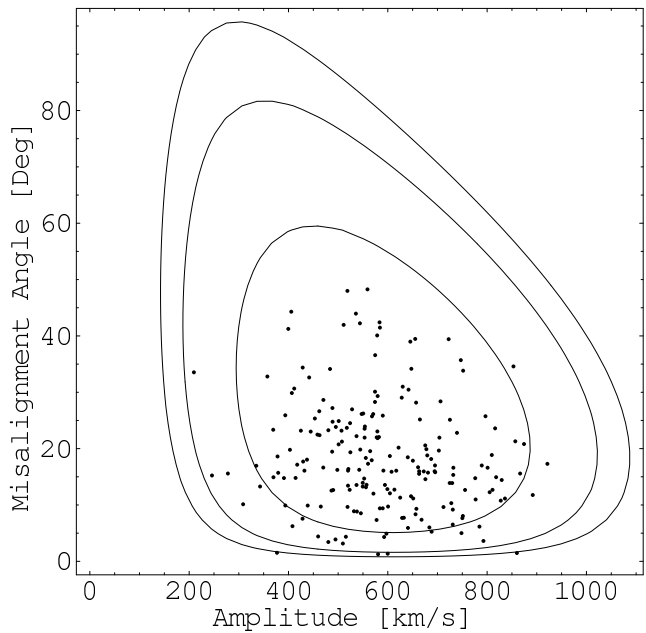}}
\caption{\label{fig:old} As in Fig.~\ref{fig:new}, but with the coherence
function given by the fit of S92.}
\end{figure}

The approach adopted by S92 implies that the finite volume effects are
also present, so they should also be modelled. We did so, and the
results are presented in Figures~\ref{fig:new}
and~\ref{fig:old}. Specifically, in both figures the scattered points
are the distribution for $g$ and $\theta$ for the LG observers,
simulated by S92 (Strauss, private communication). On this
distribution, we superimpose the probability contours resulting from
equation~(\ref{eq:dist_cond}), for observers constrained by the CMB
dipole ($v = v_{\rm CMB}$). The contours are drawn with the
finite-volume effects modelled according to S92. Figure~\ref{fig:new}
shows the contours drawn with the CF given by our
formula~(\ref{eq:func_fit}), while Figure~\ref{fig:old} shows them for
$C$ given by the fit of S92.
We see that the fit of S92 results in obviously too broad probability
contours. In contrast, our formula results in the contours which at
first look seem to be consistent with the simulated distribution. At a
closer inspection, one might worry that they are also (slightly) too
broad: e.g., outside the $95$\% probability contour there are only 5
points out of 200. However, the simulated distribution was constructed
by S92 under an additional constraint of small shear of the velocity
field around the LG, and our model does not include this. The effect
of the local shear constraint `is minor and only tightens up the
contours slightly' (S92). Still, as very little, if any,\footnote{The
discrepancy may be simply statistically insignificant.} modification
is needed, it may be just enough.

The above uncertainty has little relevance, since the aim of this
section was qualitative rather than quantitative. We showed that the
CF is important in analyses of the LG acceleration, and that an
alternative way of calibrating it, adopted by S92, also points towards
much smaller decoherence scale. We did not, however, attempt to {\em
determine\/} the CF in this way. Actually, we think that such a
determination is non-trivial, since one has to separate carefully the
influence on the correlation coefficient of nonlinear effects from
remaining effects.

In calculating Figures~\ref{fig:new} and~\ref{fig:old}, a correction
was adopted. We computed $\s_g^2$ and $\s_v^2$ according to the linear
theory [$(6 \pi^2)^{-1} \int W^2(k) P(k) {\rm d}k$]. Since on small
scales the gravity window smoothes more heavily than the velocity
window (S92), the resulting $\s_g$ was smaller than $\s_v$. This had
the effect that the predicted and simulated distributions were
slightly off-set horizontally. To correct this, we equated $\s_g$ to
$\s_v$, obtaining Figures~\ref{fig:new} and~\ref{fig:old}.

Is this slight correction valid? Nonlinear gravity is known to be
slightly larger than nonlinear velocity smoothed with the same filter
(Berlind \etal 2000, Kudlicki \etal 2001a). Therefore it may well be
that the effect of different filters compensates here, at least
partly, with the nonlinear effect. This means that in the LG
velocity--gravity comparison there are remaining residual nonlinear
effects, that still deserve further modelling. We plan to do this
elsewhere.

\section{Summary and conclusions}
\label{sec:conc}
We have studied the coherence function (CF), i.e., the
cross-correlation coefficient of the Fourier modes of the cosmic
velocity and gravity fields. This function describes nonlinear effects
in maximum-likelihood analyses of the LG acceleration. It is
important, since it affects the likelihood function and hence the
estimation of cosmological parameters (Section~\ref{sec:probab}). We
have determined the CF both analytically using perturbation theory
(Section~\ref{sec:anal}), and numerically using a hydrodynamic code
(Section~\ref{sec:num} and~\ref{sec:res}). The dependence of the
function on $\Omega_m$ and on the shape of the power spectrum has
turned out to be very weak. The only cosmological parameter that the
CF is strongly sensitive to is the normalization, $\s_8$, of the
underlying density field. We have found that the perturbative
approximation for the function is accurate as long as $\s_8$ is
smaller than about $0.3$. For higher normalizations we have provided
an analytical fit for the CF as a function of $\s_8$ and the
wavevector. The characteristic decoherence scale which our formula
predicts is an order of magnitude smaller than that found by S92.

To analyze the above discrepancy we have followed the approach of
constraining the CF adopted by S92. Specifically, we have calculated
the distribution function for the amplitude of the acceleration of the
LG and the misalignment angle, given the value of the LG velocity, and
compared with that obtained from mock {\em IRAS\/} catalogs. The
distribution resulting from our estimate of the CF turned out to be
consistent with the simulated one. In contrast, adopting S92's formula
for the function resulted in a distribution which is too broad. We
believe therefore that we have derived the correct form of the CF. The
origin of the error in the analysis of S92 remains unclear to us.

Tighter probability contours for the LG gravity imply tighter
confidence intervals for estimated cosmological parameters. The
likelihood contours for the parameters can only be drawn given the
data, thus we leave it for future analyses of the {\em observed\/} LG
acceleration. This paper strongly suggests that with proper account
for nonlinear effects in such analyses, the value of $\beta$ can be
determined with significantly greater precision\footnote{I.e., with
smaller random error.} than is currently believed.

\section*{Acknowledgments}
We thank Michael Strauss for providing the data derived from the
simulated {\em IRAS\/} acceleration for N-body observers similar to
the Local Group, and helpful suggestions. Pablo Fosalba is warmly
acknowledged for stimulating discussions. We thank the referee, Yehuda
Hofman, for his valuable comments. We are grateful to Boud Roukema for
apt comments on the text. This research has been supported in part by
the Polish State Committee for Scientific Research grants
No.~2.P03D.014.19 and 2.P03D.017.19. The numerical computations
reported here were performed at the {\em Interdisciplinary Centre for
Mathematical and Computational Modelling}, Pawi\'nskiego 5A,
PL-02-106, Warsaw, Poland.

\end{document}